\documentclass[fleqn,usenatbib]{mnras}

\usepackage{newtxtext,newtxmath}

\usepackage[T1]{fontenc}

\DeclareRobustCommand{\VAN}[3]{#2}
\let\VANthebibliography\thebibliography
\def\thebibliography{\DeclareRobustCommand{\VAN}[3]{##3}\VANthebibliography}

\usepackage{graphicx}	
\usepackage{amsmath}	
\usepackage{xcolor}
\usepackage{afterpage}
\usepackage{tikz}
\usetikzlibrary{positioning}

\newcommand{\logg} {\log \textsl{\textrm{g}}}

\newcommand{\Te} {T_{\rm eff} }

\newcommand\gta{\lower 0.5ex\hbox{$\buildrel > \over \sim\ $}} 
\newcommand\lta{\lower 0.5ex\hbox{$\buildrel < \over \sim\ $}} 

\newcommand{\Teff} {T_{\rm eff} }

\title[NPE for White Dwarf Characterization]{Neural Posterior Estimation for White Dwarf Spectroscopic Characterization}

\author[Vincent et al.]{
Olivier Vincent\thanks{E-mail: o.vincent@umontreal.ca}, 
Patrick Dufour 
and Pierre Bergeron
\\
D\'epartement de Physique, Universit\'e de Montr\'eal, C.P.~6128, Succ.~Centre-Ville, Montr\'eal, Qu\'ebec H3C 3J7, Canada\\
}

\date{Accepted XXX. Received YYY; in original form ZZZ}

\pubyear{2024}

\begin{document}
\label{firstpage}
\pagerange{\pageref{firstpage}--\pageref{lastpage}}
\maketitle

\begin{abstract}
White dwarf spectroscopic characterization is entering a big data era, with the number of spectroscopically characterized white dwarfs expected to grow from $\sim$100,000 to over 300,000 in upcoming years. Traditional methods like least-squares fitting and Markov Chain Monte Carlo have become computationally prohibitive for large-scale analysis, requiring minutes to days per star. Furthermore, these methods impose fundamental limitations on model complexity by requiring explicit likelihood functions, typically restricting them to Gaussian assumptions.
We present neural posterior estimation (NPE), a simulation-based inference technique that directly approximates posterior distributions through neural networks trained on simulated spectra. Our approach provides accurate parameter inference in milliseconds per star after upfront training costs, enabling statistical tests of the procedure's reliability. We demonstrate NPE's effectiveness on DA, DB, and carbon-atmosphere white dwarfs, validating its calibration with simulation-based calibration and tests of accuracy with random points. Application to SDSS data shows excellent agreement with previous studies, recovering parameters from previous work within 6.8\% for effective temperature and 2.1\% for surface gravity, on average. We also apply our technique on WD 1153+012, a hot DQ star with a carbon-oxygen-hydrogen atmosphere, using high-resolution spectroscopy.
This methodology combines computational efficiency with the flexibility to model complex atmospheres, making it ideal for upcoming surveys. Our approach also integrates spectroscopic and photometric constraints through an iterative procedure, providing comprehensive characterization of white dwarfs.

\end{abstract}


\section{Introduction}\label{sec:introa4}
The spectroscopic characterization of white dwarf stars aims to relate the observed spectral features to the parameters of atmospheric models, thereby constraining their physical properties. The most widely used inference methods for white dwarf parameter measurement are variants of least-squares \citep[e.g.,][]{Kepler2021, Manser2024} and Markov Chain Monte Carlo (MCMC) algorithms \citep[e.g.,][]{LopezSanjuan2022, Kiman2022}. Their sequential nature, however, makes them computationally prohibitive in scenarios involving large quantities of data or large numbers of parameters. The study of white dwarf stars is currently entering a new era of big data, recently increasing its number of spectroscopically characterized white dwarfs from $\sim$40,000 to $\sim$100,000 \citep{GarciaZamora2023, Vincent2024}, which will further be increased to over 300,000 in the upcoming years, thanks to new surveys such as SDSS-V, DESI, 4MOST and WEAVE \citep{Kollmeier2017, Cooper2023, Toloza2023, Jin2024}. Beyond spectroscopic surveys, the LSST \citep{Ivezic2019} is poised to bring photometric observations for over 156M white dwarfs in the next decade \citep{Fantin2020}. Current spectroscopic inference methods currently take up minutes to days, depending on the white dwarf type, making them impractical for large-scale application. Furthermore, least squares and MCMC methods also pose fundamental limits on the class of possible models describing the relationship between their parameters. They require explicit and tractable expression of the Jacobian/likelihood, which usually restricts them to Gaussian likelihoods, deterministic models and very few nuisance parameters. As more precise observations become available, these simplistic assumptions may not properly capture the complexity of objects such as magnetic white dwarfs \citep[e.g., the double-faced DBAH in][]{Moss2024}, D6 stars \citep{Shen2018} and cool white dwarfs with heavy metal pollution \citep{Zuckerman2007, Coutu2019}.

Recent works have explored the use of machine learning to address these issues by training neural networks models that rapidly interpolate model atmosphere grids for pure hydrogen \citep{Chandra2020} and metal-polluted \citep{BadenasAgusti2024} white dwarfs. Despite considerable speed gains, these machine learning interpolators still rely on MCMC sampling and suffer from the caveats discussed above. Outside the immediate scope of white dwarf stars, machine learning has also seen applications for the spectroscopic analysis of main-sequence stars, including neural network model interpolators \citep{Ting2019, Xiang2019}, Bayesian neural networks \citep{Leung2019} as well as various data-driven algorithms \citep{Zhang2020, Zhang2023}. Apart from the interpolators, these techniques can retrieve stellar parameters quasi-instantaneously after the upfront cost of training, but at the expense of posterior accuracy. The outputs resulting from these algorithms are either point-like estimations or probability distributions that are not true posteriors in the Bayesian sense, often enforced to follow Gaussian distributions. Data-driven techniques for white dwarf physical parameter inference are currently suboptimal due to the low number of sufficient training examples with high resolution and signal-to-noise ratio, in contrast to the 5-6M main-sequence stars training sets. It is worth mentioning that even though data-driven algorithms may not be the best approach for parameter inference, they have seen successful applications for the classification of SDSS and Gaia data \citep{GarciaZamora2023, Vincent2023, Vincent2024, Kao2024}.

Simulation-based inference \cite[SBI;][]{Cranmer2020} offers a way to both accelerate the inference process and to capture the complexity of atmospheric models required by certain white dwarfs. While SBI has seen extensive application in cosmology \citep{Cole2022, Bhardwaj2023, Saxena2023, deSanti2023} and has recently been explored for exoplanetary atmospheric retrieval \citep{Vasist2023}, it remains to be used for stellar parameter inference. In this work, we propose to use neural posterior estimation \cite[NPE;][]{Lueckmann2017, Greenberg2019}, an approach employing neural networks to directly approximate an amortized posterior distribution through simulation-based training, thereby circumventing the challenges associated with either explicit likelihood evaluations in high-dimensional parameter spaces or Bayesian inference procedures on massive observational datasets. This makes NPE fast, scalable and enables statistical tests of the procedure.

The rest of the paper is structured as follows. In section \ref{sec:methoa4}, we introduce NPE and the framework used for parameter inference. In section \ref{sec:valid}, we run several validation tests on simulated data to verify the performance of the framework. Then, in section \ref{sec:resa4}, we apply our framework to SDSS data and a carbon-atmosphere white dwarf observed at high resolution. Finally, we give concluding remarks in section \ref{sec:conca4}.

\section{Methodology}\label{sec:methoa4}

This section presents the framework used for parameter inference. Spectroscopic inference is done using a neural posterior estimation technique described in Section \ref{sec:NPE}, the data simulation procedure used to train the posterior estimation neural networks is described in section \ref{sec:simsa4}, the procedure to optionally combine constraints from photometric data in section \ref{sec:fitting} and a validation of the framework on simulated data in section \ref{sec:valid}.

\subsection{Neural Posterior Estimation}\label{sec:NPE}
In many scientific domains, complex physical processes can be accurately modeled through numerical simulations, but these simulations often lack a tractable likelihood function. Consider a stochastic simulator with parameters $\theta$ and latent variables $z$ that produces observable data $x$. The posterior distribution of interest is:

\begin{equation}
    p(\theta|x) = \frac{p(x|\theta)p(\theta)}{p(x)}
\end{equation}

\noindent
where the likelihood $p(x|\theta)$ often requires marginalizing over the latent or nuisance variables:

\begin{equation}
    p(x|\theta) = \int p(x, z|\theta) dz
\end{equation}

\noindent
and the evidence term $p(x)$ requires integration over all possible parameter values:

\begin{equation}
    p(x) = \int p(x|\theta)p(\theta) d\theta.
\end{equation}

For complex models, these integrals are typically intractable. However, even when we cannot evaluate the likelihood $p(x|\theta)$ directly, we can often bypass this issue by sampling the distribution using the forward simulator. In simulation-based approaches, an option is to use deep neural networks to parameterize universal density estimators and estimate the posterior. Neural posterior estimation is a SBI technique that directly approximates the posterior distribution $p(\theta|x)$ with a flexible density estimator. Here, we used conditional normalizing flows \citep{Papamakarios2016} providing a parameterized approximation $q_\phi(\theta|x)$ that can capture complex distributions to model the posterior. Normalizing flows are a class of models that transform a simple base distribution (e.g., a standard Gaussian) into a more complex distribution through a series of invertible transformations. Given a random variable $u$ with a simple distribution $p_u(u)$, we define a transformation $T$ such that:

\begin{equation}
    \theta = T(u) \quad \text{where} \quad u \sim p_u(u).
\end{equation}

The density of $\theta$ can then be computed using the change of variables formula:

\begin{equation}
    p_\theta(\theta) = p_u(u) \left|\det \frac{\partial T^{-1}}{\partial \theta}(\theta)\right| \quad \text{where} \quad u = T^{-1}(\theta).
\end{equation}

The key property of normalizing flows is that both $T$ and $T^{-1}$ must be differentiable and easily computable, and the Jacobian determinant must be tractable. The term "normalizing flows" arises from this process of transforming or "flowing" probability density through a sequence of invertible mappings.
To further increase their expressivity, it is possible to stack multiple simple transformations as $p_{u}=p_{u,n} \cdot p_{u,n-1} \cdot ... \cdot p_{u,1}$. 

The training process to learn the posterior involves drawing parameters from the prior $\theta \sim p(\theta)$, generating synthetic data $x \sim p(x|\theta)$ using the simulator, and then optimizing the parameters $\phi$ of the conditional flow to maximize the likelihood of the correct parameters given the generated data. This is equivalent to minimizing the Kullback-Leibler divergence between the true posterior and our approximation:

\begin{equation}
    \mathcal{L}(\phi) = \mathbb{E}_{p(x,\theta)}\left[-\log q_\phi(\theta|x)\right].
\end{equation}

Once trained, NPE provides several advantages: it allows for rapid posterior evaluations without additional simulations, it can capture non-Gaussian and multimodal posterior geometries, and it enables extremely efficient posterior sampling. This makes it a prime choice over traditional algorithms for both white dwarfs with complex atmospheres and the analysis of large samples from astronomical surveys.

More specifically, we used the APT algorithm (also known as NPE-C; \citealt{Greenberg2019}) , which improves upon previous NPE variants through a novel approach to handling the proposal distribution during training. APT can incorporate arbitrary proposal distributions while maintaining stable training, enabling more efficient exploration of the parameter space in regions of high posterior probability. The algorithm achieves this by automatically transforming between the true posterior and the proposal-adjusted posterior during training, avoiding the need for importance weights or post-hoc corrections. We used a Neural Spline Flow \citep[][]{Papamakarios2019} with 5 transformation layers, each containing rational-quadratic splines with 20 bins that progressively transform a standard normal distribution into the final posterior. Each transformation is controlled by a neural network with 45 hidden units per layer that conditions on the observation data, which is normalized using structured standardization.

Once a posterior distribution has been obtained for a given observation, we determine the best-fit parameter values using Maximum A Posteriori (MAP) estimation, which identifies the point of highest probability density. Unlike simple mean or median calculations, this approach respects the potentially complex, non-Gaussian shape of the posterior. To estimate the MAP, we first apply a Kernel Density Estimate (KDE) to the posterior samples using Scott's rule for bandwidth selection implemented in \texttt{scikit-learn} \citep{Pedregosa2011}. We then find the density peak using negative-likehood minimization with the L-BFGS-B algorithm implemented in \texttt{scipy} \citep{Virtanen2020}. We quantify parameter uncertainties using 16th and 84th percentile-based credible intervals from the posterior samples.

\subsection{Spectroscopic Data Simulations and Training}\label{sec:simsa4}
We trained three distinct neural posterior estimators using three distinct simulated training sets: one for DA and one for DB white dwarfs observed by the SDSS, and one for a Hot DQ observed at the Multiple Mirror Telescope Observatory (MMTO). The simulations were generated using model atmospheres explained in detail in Section 4.2 of \citet{Vincent2024}. We used the models in \citet[DA;][]{Blouin2018, Tremblay2011, Bedard2020} for pure hydrogen (DA), the models by \citet{Blouin2019,GenestBeaulieu2019,Bedard2020} for pure helium (DB) and \citet{Dufour2005, Dufour2008} with updated stark broadening \citep{Dufour2011} for hot carbon-oxygen (Hot DQ) atmospheres. A summary of the grids, parameter ranges and simulation details is provided in Table \ref{tab:gridsa4}. The grids were made with $\Teff$ steps of 1000~K for DA/DB and 500~K for Hot DQ, and steps of 0.5~dex in $\logg$ and all other abundance ratios.

\begin{table}
    \centering
    \begin{tabular}{|p{1.1cm}|p{2.7cm}|p{1.15cm}|p{1.3cm}|}
         \hline
         \hline
         \centering{Grid} & Parameter ranges & Number of simulations & Spectral region \\
         \hline
         \hline
         \centering{Pure H (DA)} & 
         $\Teff =$ [1500, 150,000] \newline
         $\logg =$ [6.5, 9.5] & 
         30,000 & Balmer lines\\
         \hline
         \centering{Pure He (DB)} & 
         $\Teff =$ [11,000, 150,000] \newline
         $\logg =$ [6.5, 9.5] & 
         30,000 & 3842-7000~\r{A}\\
         \hline
         \centering{Carbon (Hot DQ)} & 
         $\Teff =$ [12,000, 25,000] \newline
         $\logg =$ [7, 9.2] \newline
         log {\rm C}/{\rm H} = [$-$4, 4] \newline
         log O/C = [0, 1] & 
         40,000 & 3500-5200~\r{A} \\
         \hline
         \hline
    \end{tabular}
    \caption{Overview of the synthetic spectra generated for the training set.}
    \label{tab:gridsa4}
\end{table}

To simulate the noise and resolution effects of spectroscopic observations, we followed the same procedure as \citet{Vincent2025}. Briefly, we replicated DA and DB observations from the SDSS by first linearly interpolating pristine spectra from the model grids and convolving them with a Gaussian kernel with a full width at half maximum (FWHM) of 3~\r{A} and interpolating the flux to match the wavelength sampling between 3842 and 7000~\r{A}. We applied random Gaussian noise to each pixel\footnote{In this work, a pixel refers to a single wavelength-flux.} to achieve a median signal-to-noise ratio (SNR) ranging between 9 and 50 for each spectrum and applied a random radial velocity shift drawn from a uniform distribution between $-$100 and 100 km~s$^{-1}$. For the MMT simulations, we use a FWHM of 1~\r{A}, interpolate the flux to match the observed wavelength grid between 3500 and 6000~\r{A}, use a median SNR ranging from 20 to 40 and apply the same radial velocity shifts.

Once the spectra have been simulated, we normalized their continuum to unity and sliced out certain regions for the fitting procedure. For DA spectra, we closely follow the procedure proposed by \citet{Chandra2020}, which is inspired from \citet{Bergeron1995} and subsequent work. We first selected a region around each Balmer line from H$\alpha$ to H8, fitted a Voigt profile added to a linear function to each region, and normalized them using the linear component of the fit. More specifically, the regions are selected by cropping out parts of the spectra at fixed distance (in \r{A}) from the center of each Balmer line: 300 for H$\alpha$, 200 for H$\beta$, 120 for H$\gamma$, 75 for H$\delta$, 55 for H$\epsilon$, and 30 for H8. Other regions from the spectra are discarded.

\begin{figure}
    \centering
    \includegraphics[width=1.\linewidth]{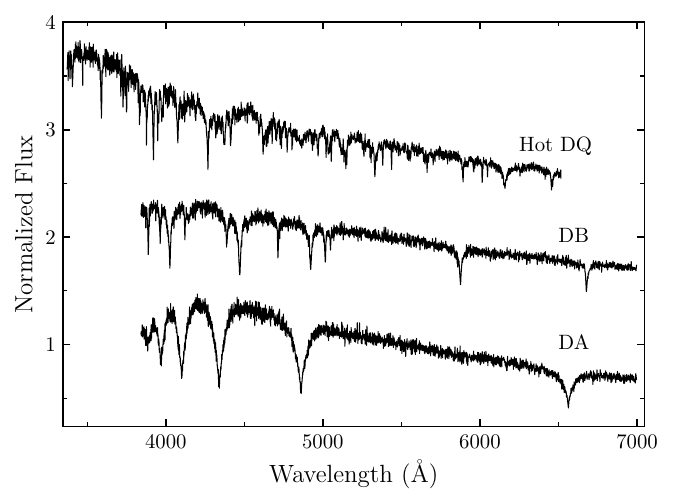}
    \caption{Examples of unnormalized simulated spectra.}
    \label{fig:examples}
\end{figure}

For DB and Hot DQ spectra, we first used a running median filter with a window size of 150~\r{A} to estimate the underlying continuum and then applied Gaussian smoothing with a $\sigma=10$~\r{A} to the estimated continuum to remove small-scale variations. The original spectra are divided by the smoothed continuum to produce the continuum-normalized spectra. For Hot DQ spectra, we retained only the 3500-5300~\r{A} region for our fitting procedure, while for DB spectra we kept the full 3842-7000~\r{A} region.

We generated 30,000 unique simulations for DA and DB white dwarfs and 200,000 simulations for carbon atmospheres by uniformly sampling the physical parameter listed on Table \ref{tab:gridsa4}. The NPE for DA and DBs is trained on the entirety of their respective simulation sets, while the NPE is trained on a random selection of 40,000 simulations. The remaining 160,000 are archived to rapidly train new NPE with constraints on $\logg$ from photometric data during the spectrophotometric fitting procedure, explained in the next section. An additional 10,000 spectra are simulated for each atmosphere type to validate the performance of the NPE during training.

\subsection{Spectrophotometric Fitting}\label{sec:fitting}
Spectroscopic and photometric data can be combined to provide complementary constraints to infer the physical parameters of white dwarf stars. Spectroscopic data allows for measurements of effective temperature ($\Teff$), surface gravity ($\logg$), and chemical abundances $Z$ through detailed modeling of line profiles. In standard spectroscopic fitting procedures, including our approach, the stellar continuum contribution is typically excluded either by flattening the spectral energy distribution through various normalization techniques \citep[e.g.,][]{Bergeron1992, Bergeron2011} or by fitting a polynomial concurrently to account for missing physics or flux calibration effects \citep[e.g.,][]{Dufour2005, Dufour2007}. Conversely, photometric analysis captures the global morphology of the spectral energy distribution by modeling broadband-averaged fluxes across an extensive wavelength range.

While the combination of the two data types can be used to iteratively refine the parameter constraints \citep[e.g.,][]{Dufour2005, Dufour2007} or to resolve degeneracies in line broadening effects such as the symmetrical "hot" and "cold" $\Teff$ solutions in DA white dwarfs \citep[see][and references therein]{Tremblay2011}, their integration requires careful consideration. A well-documented systematic offset exists between parameters derived from spectroscopic versus photometric analyses \citep{Bergeron2019, Tremblay2019b}, and substantial discrepancies may indicate unresolved white dwarf binary systems \citep{GenestBeaulieu2019}. A naive combination of probability distributions from both data types could therefore introduce biases in the final parameter solutions.

In this work, we applied our combined analysis procedure specifically to carbon-atmosphere white dwarfs. The uncertainties in carbon and oxygen line broadening physics render $\logg$ measurements from spectroscopy alone unreliable, necessitating additional constraints from photometry. For DB white dwarfs, we do not incorporate photometric data in parameter inference, while for DA white dwarfs, photometric data serves only to select the correct $\Teff$ from the spectroscopic posterior distributions. Our procedure for carbon-atmosphere white dwarfs is as follows.

We first obtain the spectroscopic posterior distribution using the NPE for all parameters. {For the photometric fit, we follow a similar approach to the technique described in \cite{Bergeron1997}. Briefly, the standardized synthetic SDSS magnitudes are converted into average fluxes using appropriate zero-points and conversion equations \citep{Holberg2006}. Model photometry is then calculated for class-specific model grids by integrating the monochromatic Eddington fluxes over each bandpass. These model fluxes depend on the effective temperature $T_\mathrm{eff}$, the surface gravity $\logg$ and chemical composition. The observed and model fluxes are then related to each other via the solid angle $\pi(R/D)^2$, where $R$ is the stellar radius and $D$ is the distance from Earth. Since the distance is known from Gaia parallax measurements, the radius can be measured directly and converted into stellar mass $M$ using evolutionary models, which provide a temperature-dependent mass-radius relation. We rely on the evolutionary models described
in \cite{Bedard2020} with C/O cores, $q({\rm He})\equiv \log M_{\rm
  He}/M_{\star}=10^{-2}$ and $q({\rm H})=10^{-4}$, which are
representative of H-atmosphere white dwarfs, and $q({\rm He})=10^{-2}$
and $q({\rm H})=10^{-10}$, which are representative of He-atmosphere
white dwarfs\footnote{The models can be found here: \url{https://www.astro.umontreal.ca/~bergeron/CoolingModels}}.
Since photometric inference represents a lower-dimensional problem, we employ traditional Markov Chain Monte Carlo (MCMC) sampling using the \texttt{emcee} library \citep{ForemanMackey2013} to infer $\Te$, $\logg$ and the solid angle, $\pi(R/D)^2$, where $R$ is the radius of the star, and $D$ its distance from Earth which can be obtained directly from Gaia parallax.} This approach incorporates the spectroscopic posteriors for chemical abundances as prior distributions. To ensure proper uncertainty propagation, we marginalized over chemical abundances rather than treating them as free parameters. Our likelihood function for $N$ observed magnitudes assumes a Gaussian distribution:

\begin{equation}
    \log \mathcal{L}_{\mathrm{phot}} \propto \sum_{i=1}^{N} \frac{(m_{i,\mathrm{obs}} - m_{i,\mathrm{model}})^2}{\sigma_i^2}~,
\end{equation}

\noindent where $m_{i,\mathrm{obs}}$ is the observed magnitude in photometric filter $i$, $m_{i,\mathrm{model}}$ is the corresponding synthetic magnitude predicted by our model, and $\sigma_i$ is the measurement uncertainty associated with the observed magnitude. {Photometric catalogs are known to have reported uncertainties that are under-estimated, since they do not adequately factor in zeropoint errors. we follow \cite{Bergeron2019} and adopt a lower limit of 0.03 mag uncertainty in all bandpasses.} We also included a likelihood term to infer the distance of the star if a parallax measurement is available:

\begin{equation}
    \log \mathcal{L}_{\mathrm{D}} \propto \left(\frac{\pi - 1000/D}{\sigma_\pi}\right)^2~,
\end{equation}

\noindent where $\pi$ is the observed parallax in milliarcseconds, $D$ is the distance in parsecs, and $\sigma_\pi$ is the uncertainty in the parallax measurement. Furthermore, we included the physically motivated distance prior proposed by \citep{BailerJones2018}, $p(D) \propto D^2 \exp(-D/L)$ representing a geometric volume with scale length $L$ set to $L=100$~pc here. This results in the following photometric posterior:

\begin{equation} \label{eq1}
\begin{split}
p(\Teff, \logg, D | x_\mathrm{phot}, Z_\mathrm{spec}, \pi) \propto  & \mathcal{L}_{\text{phot}}(\Teff, \logg)  \times \mathcal{L}_{\text{D}}(\pi) \\
&  \times p(\Teff, \logg |Z_\mathrm{spec})\times p(D) ~.
\end{split}
\end{equation}

To model $p(\Teff, \logg |Z_\mathrm{spec})$, we fitted a Kernel Density Estimator on the spectroscopic posterior distribution with a Gaussian kernel and Scott's bandwidth. We run inference with 32 walkers, 1000 burn-in steps, and 2500 sampling steps.

Once the photometric posterior has been obtained, we created a prior distribution for the next spectroscopic inference step by fitting a Gaussian on the marginal $\logg$ distribution. The Gaussian is truncated at the boundaries of the model atmosphere grid. The priors for other parameters remain uniform. We sampled from archived simulations to form new training set based on these priors. If we have less than 10,000 simulations, new ones are generated until we reach this number. A new NPE is trained by excluding $\logg$ as a free parameter. We instead treated it as a nuisance parameter and marginalized over the distribution obtained from photometry. We then repeated the photometric and spectroscopic (with marginalized $\logg$) inferences until $\logg$ does not change more than 0.1 dex between iterations. This usually happens after 2 or 3 iterations. 

To summarize, we implemented an iterative approach that begins with spectroscopic inference using NPE to obtain posterior distributions for all physical parameters. Next, we performed photometric inference through MCMC sampling, incorporating the spectroscopic posteriors as nuisance parameter priors for chemical abundance parameters. From the resulting photometric posterior, we derived a Gaussian prior for $\logg$ that constrains the subsequent spectroscopic inference step. We then repeated this cycle, alternating between photometric and spectroscopic inference with marginalized $\logg$, until convergence is achieved (defined as $\logg$ changing by less than 0.1 dex between iterations).

\section{Posterior Calibration}\label{sec:valid}
\begin{figure*}
    \centering
    \includegraphics[width=0.99\linewidth]{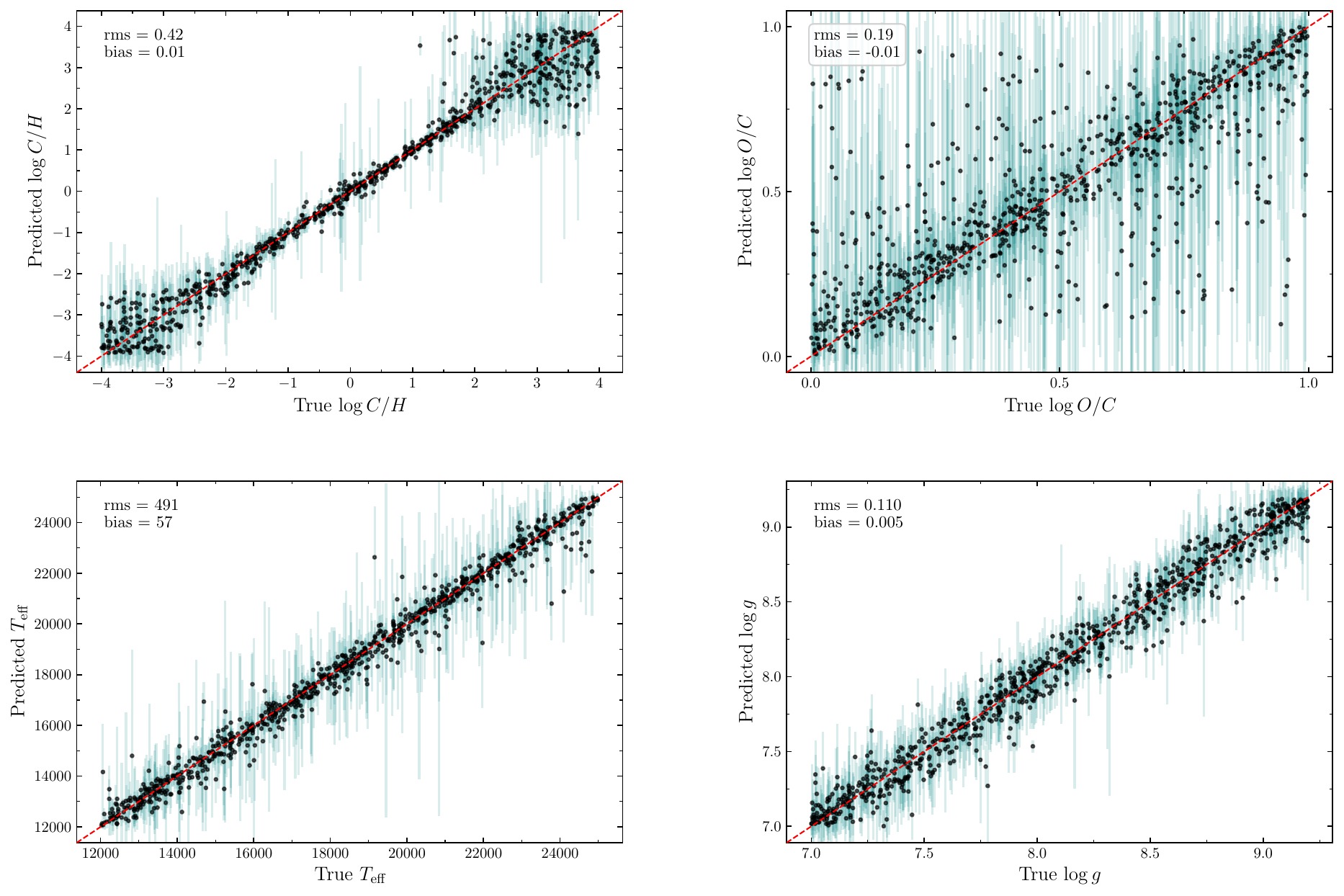}
    \caption[Parameter recovery test using our neural posterior estimation for carbon-atmosphere simulated spectra]{Parameter recovery test using our neural posterior estimation for 2000 simulated carbon-atmosphere spectra between SNR 20 and 40.}
    \label{fig:recovDQ}
\end{figure*}

\begin{figure*}
    \centering
    \includegraphics[width=0.84\linewidth]{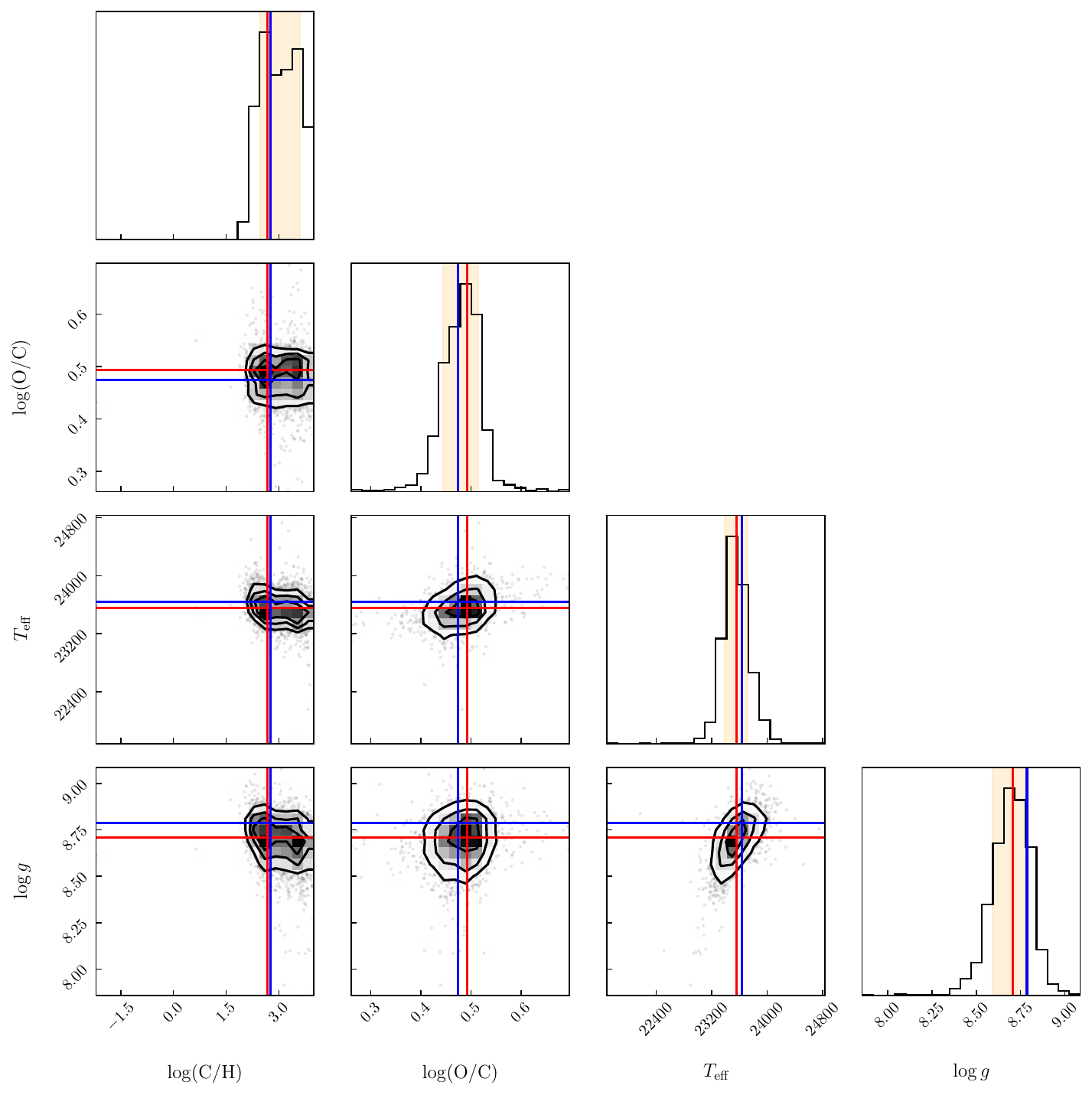}
    \includegraphics[width=0.8\linewidth]{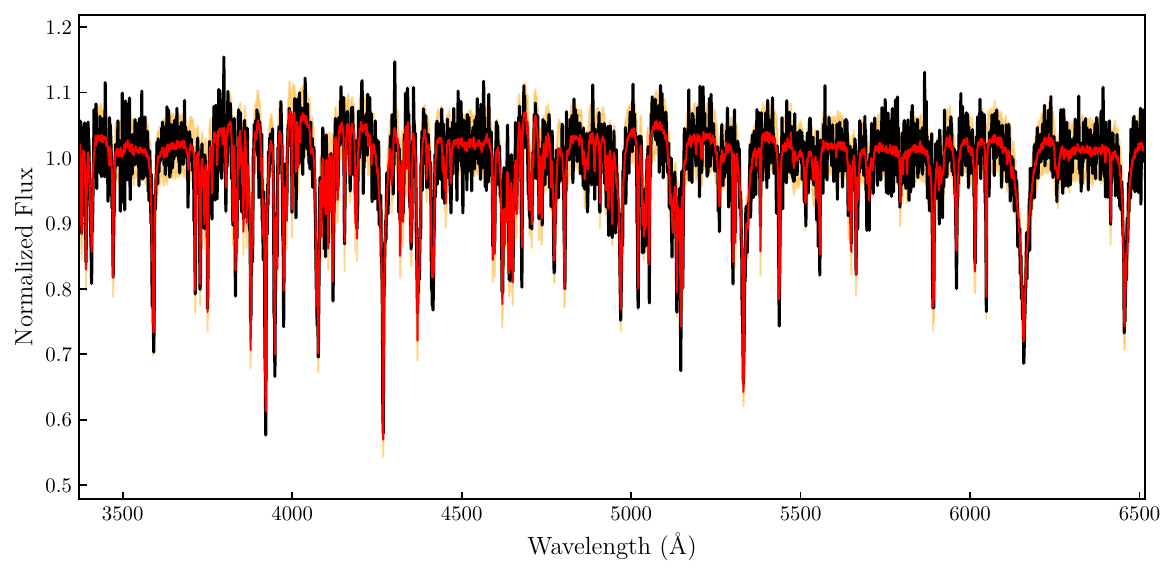}
    \caption[Fit example on one of our simulated carbon-atmosphere objects]{Fit example on one of our simulated carbon-atmosphere objects. The top panel shows the corner plot, with true parameters indicating in blue and best-fit parameters in red. In the bottom panel, the continuum-normalized simulation is shown, overlaid with the average of the top 100 highest probability samples (red) from the posterior and their standard deviation (orange contour).}
    \label{fig:DQtestfit}
\end{figure*}

In this section, we assess the validity of our neural posterior inference methodology on spectroscopic simulations of carbon-atmosphere white dwarfs. To this end, we simulate 2000 new spectra and run the procedure described in section \ref{sec:NPE}. The recovered parameters are shown in Figure \ref{fig:recovDQ} and an example fit is shown in Figure \ref{fig:DQtestfit} for one of our simulated objects. Overall, the parameters are properly recovered, with no significant bias in measured values. The scatter (calculated as the root mean square here) for $\Teff$ and $\logg$ is a reflection of the precision on the measured parameters when analyzing spectra with SNR ranging between 20 and 40. An increase in scatter for $\log {\rm C}/{\rm H}$ values approximately lower than $-2$ or greater than 2 indicates that the effect of increasing or decreasing the hydrogen content does not significantly change spectral features between 3500 and 5000~\r{A}. The scatter is most notable for the ${\rm O}/{\rm C}$ abundances which can be explained by the disappearance of the oxygen triplet around 3700~\r{A} at temperatures below 18,000~K, making oxygen difficult to constrain. 

Next, we verify the calibration of the spectroscopic posteriors with a series of tests. These tests ensure we obtain posterior distributions that are well-calibrated, meaning there are no systematic biases in the predictions and the uncertainties are neither over- nor under-confident. 
Simulation-based calibration \citep{Talts2020} provides several measures to check whether the variances of the posterior are balanced. The basic ideas behind SBC are: (1) ranks of ground truth parameters under the inferred posterior samples follow a uniform distribution and (2) samples from the data-averaged posterior (ensemble of randomly chosen posterior samples given multiple distinct observations) are distributed according to the prior. As such, SBC can be viewed as a necessary condition, but not sufficient, for a valid inference algorithm. 

The theoretical justification behind uniform rank distributions in SBC stems from a key property of Bayesian joint distributions. As shown in \citet{Talts2020}, when we sample $\tilde{\theta} \sim \pi(\theta)$ from the prior, generate data $\tilde{y} \sim \pi(y|\tilde{\theta})$, and then sample $\{\theta_1,...,\theta_L\} \sim \pi(\theta|\tilde{y})$ from the posterior, the rank statistic of $\tilde{\theta}$ relative to posterior samples will be uniformly distributed across integers $[0,L]$ for any one-dimensional function $f: \Theta \rightarrow \mathbb{R}$. This uniformity emerges because the joint distribution $\pi(\theta,y)$ ensures that when marginalizing over all possible datasets, the data-averaged posterior equals the prior: 

\begin{equation}\label{eq:dap}
    \pi(\theta) = \int d\tilde{y}d\tilde{\theta}\pi(\theta|\tilde{y})\pi(\tilde{y}|\tilde{\theta})\pi(\tilde{\theta})~.
\end{equation}

\noindent Consequently, the probability mass function of the rank statistic can be reduced to $\pi(r) = \frac{1}{L+1}$ through change of variables and properties of beta functions.

Following the recommendations in \citet{Talts2020}, we randomly select 1000 samples from 150 posteriors among our 2000 carbon-atmosphere simulations, sort them and rank each parameter separately. We run a Kolmogorov-Smirnov (KS) test to calculate p-values on the null hypothesis that the samples from ranks are drawn from a uniform distribution. The null hypothesis is usually rejected if the p-values fall below a significance threshold of 0.05. We obtain p-values between 0.2 and 0.9 for every parameter, indicating that the parameter ranks are close to an uniform distribution. We then compute the data-averaged posterior (DAP, see equation \ref{eq:dap}) by randomly selecting 1000 samples from the same 150 posteriors as above. We run a Classifier Two-Sample Test (C2ST) following \citet{LopezPaz2018} and \citet{Friedman2004} by training a Random Forest binary classifier using the default implementation from \texttt{scikit-learn} to distinguish between the DAP and prior (uniform in this case) distributions. On average, we obtain probabilities between 0.45 and 0.6 for all parameters, indicating most of the DAP is properly uniform, but some parameter regions may be slightly miscalibrated.
       
We complement SBC with Tests of Accuracy with Random Points \citep[TARP;][]{Lemos2023}, which verifies posterior accuracy by checking if the expected coverage probability equals the nominal value ($1-\alpha$) for all credibility levels $\alpha$ when using these randomly-positioned regions. Briefly, TARP works by aggregating posterior samples from a set of parameter-simulation pairs $(\theta, \tilde{y})$, calculating the distance between the samples and a set of reference points $\theta_\mathrm{r}$, and then estimating the coverage by counting for how many of the posterior samples are within a certain distance threshold. In contrast to SBC, TARP provides a necessary and sufficient condition for posterior accuracy. By creating credible regions centered at random reference points, rather than around high-density areas, and measuring whether true parameters fall within them at the correct rate, TARP effectively checks the accuracy of the posterior everywhere. Here, we randomly select 1000 samples from 200 posteriors among our 5200 carbon-atmosphere simulations and uniformly sample the grid parameter range for reference points. We show our expected coverage curve for the three atmosphere types in Figure \ref{fig:TARP}. Significant deviations from the ideal case (dashed black line) would indicate miscalibration issues (see Figure 2 of \citet{Lemos2023} for an illustration). For all three atmosphere types, the TARP curves indicate the posteriors are very close to being perfectly calibrated with a slight tendency towards under-confidence.

\begin{figure}
    \centering
    \includegraphics[width=1\linewidth]{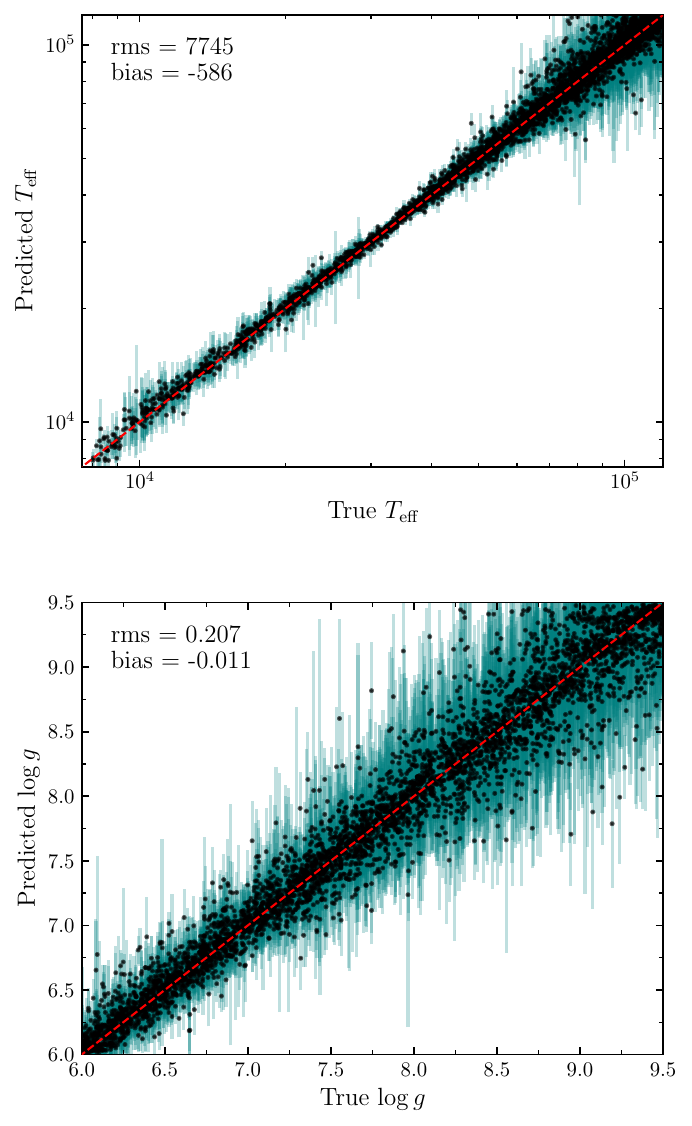}
    \caption[Parameter recovery test using our neural posterior estimation for simulated SDSS DA spectra]{Parameter recovery test using our neural posterior estimation for 5000 simulated SDSS DA spectra between SNR 9 and 50.}
    \label{fig:DArecov}
\end{figure}

\begin{figure}
    \centering
    \includegraphics[width=1\linewidth]{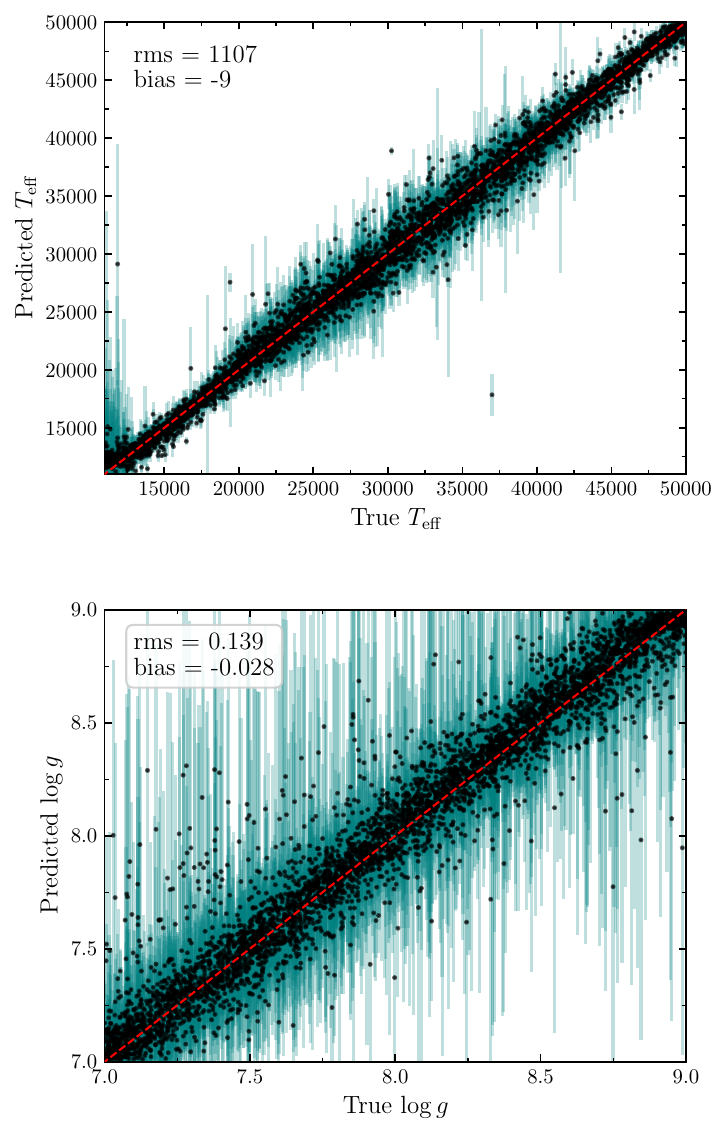}
    \caption[Parameter recovery test using our neural posterior estimation for simulated SDSS DB spectra]{Parameter recovery test using our neural posterior estimation for 5000 simulated SDSS DB spectra between SNR 9 and 50.}
    \label{fig:DBrecov}
\end{figure}

\begin{figure}
    \centering
    \includegraphics[width=1\linewidth]{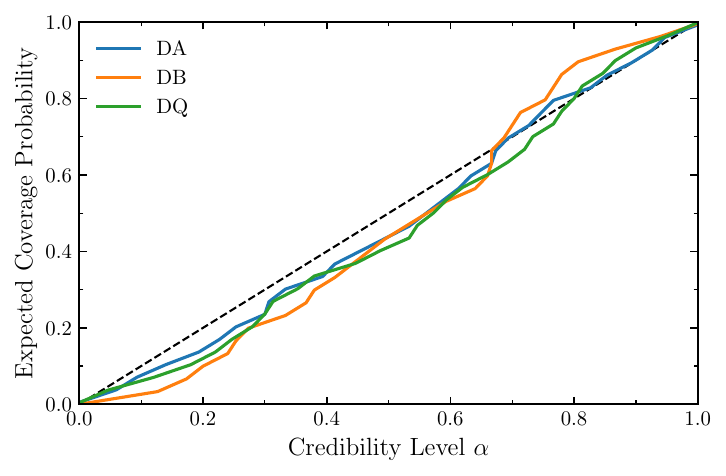}
    \caption[Tests of Accuracy with Random Points curves for our three atmosphere types]{Tests of Accuracy with Random Points (TARP) curves for our three atmosphere types. Perfectly calibrated posterior estimators will follow the black dashed line, while curves either above or below it indicate over- and under-confident posteriors, respectively. In our case, the estimators are close to being perfectly calibrated with a slight tendency towards under-confidence.}
    \label{fig:TARP}
\end{figure}

Overall, NPE provides accurate and well-calibrated results. We also ran the same suite of tests on DA and DB atmospheres, which are simpler cases as they only require two parameters, and found the same conclusions. Figure \ref{fig:DArecov} shows the parameter recovery test for simulated SDSS DA white dwarfs, while results for simulated SDSS DB white dwarfs are shown in Figure \ref{fig:DBrecov}. We find no significant systematic bias in our recovered parameters, but we do note an increase in scatter that appears to depend on temperature in both cases. For DAs, the scatter increases with temperature around $\Teff=40,000$~K, which can be attributed to a loss of sensitivity in optical observations at higher temperatures \citep{Bedard2020}. As white dwarfs become hotter, most of their radiation shifts toward the ultraviolet, leaving optical wavelengths sampling primarily the Rayleigh-Jeans tail of the spectral energy distribution. Spectroscopically, hydrogen line profiles become progressively less sensitive to atmospheric parameters with increasing temperature. In the case of DB white dwarfs, the increased scatter between 20,000 and 30,000~K is already discussed in section \ref{sec:resa4}. However, our results suggest that the decreased sensitivity of optical helium line profiles to $\Teff$ and $\logg$ extends to $\sim$45,000~K rather than to 30,000~K as discussed in \citet{Bergeron2011}, at least in the case of SDSS optical spectroscopy. 

\section{Application to SDSS and MMTO Data}\label{sec:resa4}
\citet{Tremblay2019} derived spectroscopic $\Teff$ and $\logg$ for a sample of 5327 SDSS DA white dwarfs and obtained values consistent with results obtained by other groups and photometry, with typical root mean
square (rms) discrepancy between stellar labels derived by different methods being on the order of 500 K in $\Teff$ and 0.1 dex in $\logg$. The Tremblay et al. study was also used as a comparison point for the different machine learning pipelines in \citet{Chandra2020}. Here, we also use the stellar labels from \citet{Tremblay2019} as a "ground truth" to test the methods described in this work.

\begin{figure}
    \centering
    \includegraphics[width=0.99\linewidth]{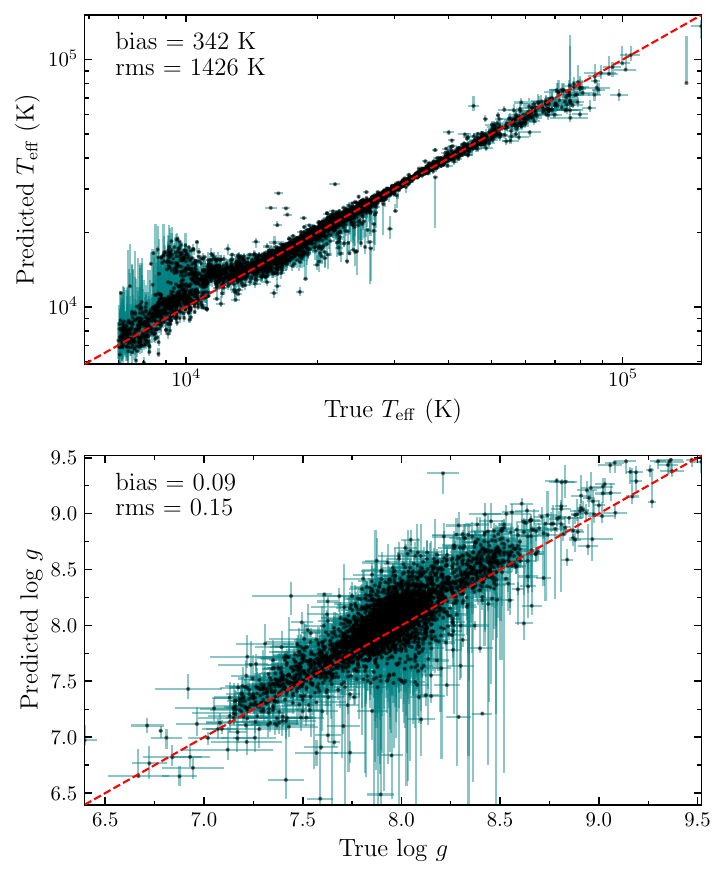}
    \caption[Comparison of parameters for SDSS DA white dwarfs obtained by our method and those of Tremblay et al. (2019)]{Comparison of parameters for 5327 SDSS DA white dwarfs obtained by our method and those of \citet{Tremblay2019b}. Also shown is the the bias and spread for objects with $\Teff$ between 7000 and 40,000~K.}
    \label{fig:compDA}
\end{figure}

We run our fitting procedure on all 5327 spectra, obtaining the full posterior distribution in the order of milliseconds per star on average. To break the "cold" and "hot" solution degeneracy, we use the reference $\Teff$ value as a seed to select the appropriate solution in multimodal posteriors. In practice, this is usually done by measuring $\Teff$ using photometric data. {The approximation used here is sufficient, as a self-consistent fit would yield identical results}. We present the comparison between our labels and those from \citet{Tremblay2019} in Figure \ref{fig:compDA}. The results are shown in Figure \ref{fig:compDA}. Our technique recovers $\Teff$ within 6.1\% and $\logg$ within 1.6\% on average across a temperature range of 7000 to 150,000~K. The most significant offsets happen near the hot/cold degeneracy point near 12,000~K and at $\Teff>70,000$~K where the broadening of hydrogen lines becomes less sensitive to these parameters \citep{Bedard2020}. {The normalizing flow, which approximates the posterior distribution, struggles to separate "hot" and "cold" modes when they are very close (e.g., near $\Teff \sim12,000$~K). We believe this issue, observed in a few spectra, creates small offsets that ultimately result in a higher RMS.} Between 7000 and 40,000~K, we obtain a bias of 339~K and 0.09~dex between predictions and true values with a spread of 1420~K and 0.15~dex. These results are comparable with the performance of the generative neural network trained on SDSS DA observations by \citet{Chandra2020}, which had bias of 265~K and 0.01~dex and a spread of 801~K and 0.12~dex on $\Teff$ and $\logg$, respectively, for stars between 7000 and 40,000~K.

\begin{figure}
    \centering
    \includegraphics[width=0.99\linewidth]{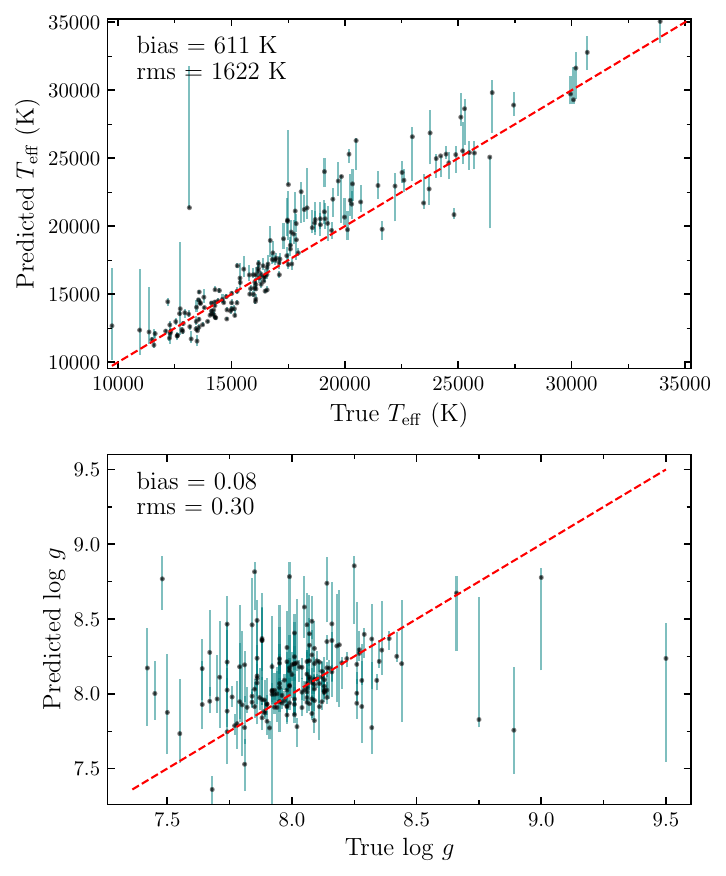}
    \caption[Comparison of parameters for SDSS DB white dwarfs obtained by our method and those in Genest-Beaulieu et al. (2019). The bias and spread for all objects are shown.]{Comparison of parameters for 300 SDSS DB white dwarfs obtained by our method and those in \citet{GenestBeaulieu2019}. The bias and spread for all objects are shown.}
    \label{fig:compDB}
\end{figure}

We now test our method on DB white dwarfs from \citet{GenestBeaulieu2019}, who derived the physical properties of 1915 white dwarfs with helium-dominated atmospheres. We select 300 stars with the DB classification within their work to exclude any object with additional traces. We run our spectroscopic fitting procedure on this sample, again obtaining the full posterior distribution in the order of milliseconds per star on average despite fitting the full optical spectrum, or $\sim$3 times more pixels than the DA procedure for which only regions around Balmer lines were used. If a posterior is detected to be multimodal, we select the solution with the closest $\Teff$ to the value from \citet{GenestBeaulieu2019}. The results are shown in Figure \ref{fig:compDB}. Our technique recovers $\Teff$ within 7.6\% and $\logg$ within 2.6\% on average. The scatter is slightly higher between 18,000 and 30,000~K, a temperature range where optical spectroscopy for helium-atmosphere white dwarfs is essentially constant \citep{Bergeron2011}. Briefly, the lack of change in optical spectral features is because as temperature increases, more helium atoms are excited to higher energy states (increasing line strength), but at the same time, more helium atoms become ionized (decreasing line strength). These opposing effects balance each other in this temperature range, resulting in nearly identical spectral appearances despite significant differences in effective temperature. We also note that small offsets due to our continuum normalization technique, based on running median filtering, being different from the one used by \citet{GenestBeaulieu2019}, who identify the continuum pixels using spectral regions known to have no helium features. {The determination of $\log g$ for DB stars is intrinsically more uncertain. As discussed in \citet{GenestBeaulieu2019}, the analysis of DB white dwarfs remains more challenging and less well understood than for DA stars, and these known issues are reflected in our results, particularly in Figure \ref{fig:compDB}.}

We now turn to the analysis of WD 1153+012 (SDSS J115305+005646), first identified as a hot DQ white dwarf in \citet{Dufour2008}, who reported preliminary parameters of $\Te = 21,650$~K from SDSS spectroscopy and 21,400~K from $ugriz$ photometry, both assuming a fixed value of $\logg=8.0$. They did not provide quantitative abundance ratios, and oxygen was believed to be at least an order of magnitude less abundant than carbon based on the absence of clear oxygen features in the SDSS data. Subsequent high SNR spectroscopy obtained with the MMT revealed weak oxygen lines \citep{Dufour2009}, suggesting potentially significant oxygen abundance, but no comprehensive multi-element analysis has been performed until now.

\begin{figure*}
    \centering
    \includegraphics[width=0.99\linewidth]{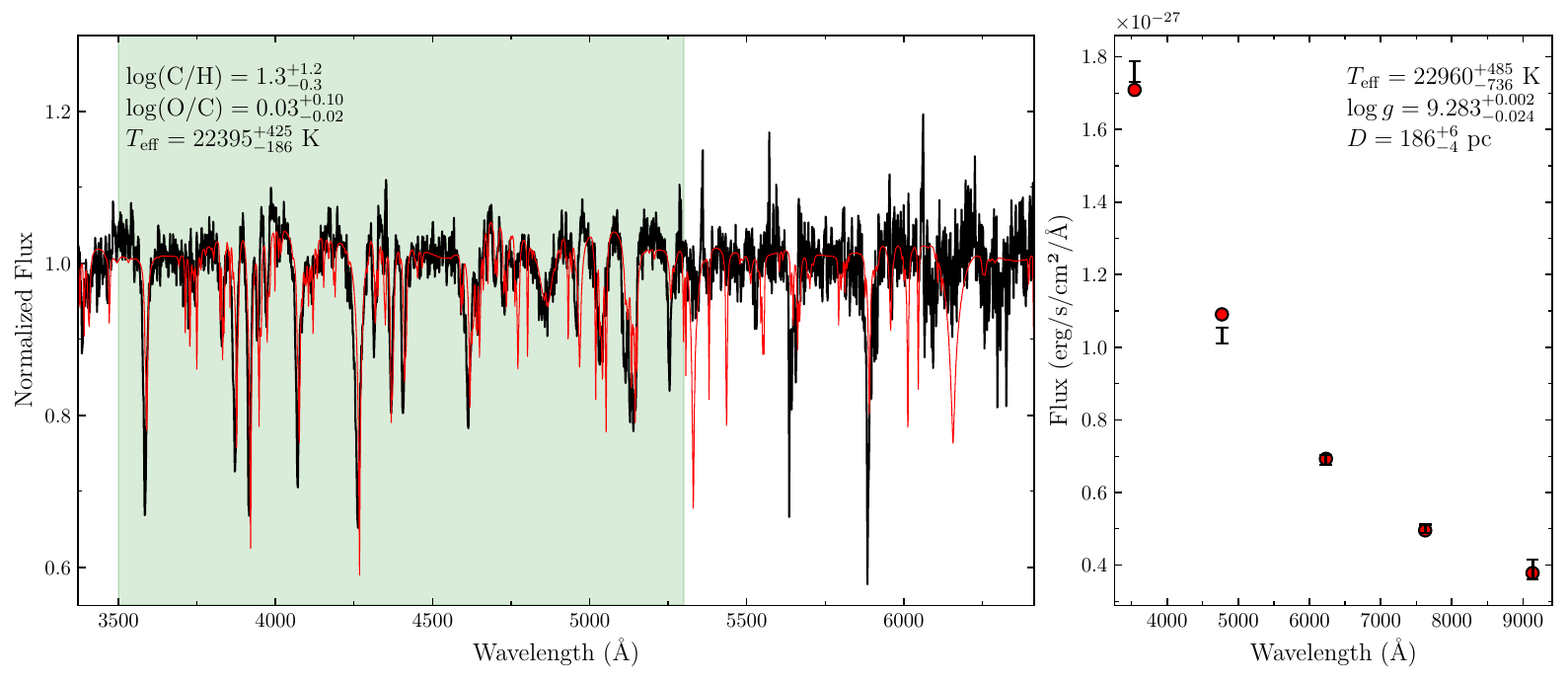}
    \caption[Observation and spectrophotometric best-fit solution to the oxygen-carbon atmosphere white dwarf WD 1153+012]{Observation and spectrophotometric best-fit solution to the oxygen-carbon atmosphere white dwarf WD 1153+012. Left side: Continuum-normalized MMTO observation (black) and best solution (red) using the spectroscopic parameters indicated at the top left in combination with the surface gravity measured from photometry. The green region indicates the portion of the spectrum used for fitting. Right: SDSS $ugriz$ photometry (black error bars) and best fit solution (red) using the photemetric parameters indicated on the top right in combination with the chemical abundances measured from spectroscopy.}
    \label{fig:1153}
\end{figure*}

Unlike the analysis of the DA and DB samples, we also incorporate photometric data in the fitting procedure in order to further constrain $\logg$ (see Section \ref{sec:fitting}). The resulting spectroscopic and photometric fits are shown in Figure \ref{fig:1153}. 
Our neural posterior estimation yields a consistent solution across both spectroscopic and photometric data. The spectroscopic data favor $\Teff = 22,395^{+425}_{-186}$~K while the photometric analysis indicates $\Teff = 22,960^{+485}_{-736}$~K. This represents a difference of approximately 600 K, which falls within the combined uncertainties. The asymmetric posterior distribution suggests that the true effective temperature likely falls between these values, slightly higher than the temperature reported in \citet{Dufour2008}.
We obtain a surface gravity of $\logg = 9.28^{+0.002}_{-0.024}$, also higher than the previously fixed value of 8.0 in \citet{Dufour2008}. The upper uncertainty on $\logg$ is limited by our model grid boundary at $\logg = 9.2$ and the true value may be even higher. We note that the initial abundance measurements could be partially influenced by magnetic broadening effects, as hot DQs are typically found to be magnetic \citep{Dufour2013}, potentially leading to an overestimation of surface gravity if magnetic effects are not fully accounted for in the model atmospheres.

For the atmospheric composition, we obtain $\log {\rm C}/{\rm H} = 1.3^{+1.2}_{-0.3}$ indicating that increasing the amount of carbon does not have a visible effect on the spectrum between 3500-5300~\r{A}, and is hence a lower limit on the abundance. 
The oxygen abundance measurement represents a significant advancement over previous studies, with our analysis yielding $\log O/C = 0.03^{+0.10}_{-0.02}$. This value approaches the lower boundary of our model grid (log O/C = 0), suggesting that carbon could be slightly more abundant than oxygen by mass. The slightly deeper predicted oxygen triplet at 3700~\r{A} compared to the observed spectrum suggests that the true O/C ratio may be marginally lower than our best-fit value, but oxygen remains a significant constituent of the atmosphere. This finding aligns with preliminary suggestions from \citet{Dufour2011} that hot DQs might have oxygen abundances as high as 50\% in some cases.

The entire fitting procedure for this complex object took approximately two minutes, including over one minute for the photometric MCMC, 30 seconds for training new neural posterior estimators without $\logg$ as a free parameter, and 5 seconds for automatic mode identification. The actual spectroscopic posterior sampling time remained negligible, highlighting the efficiency of our neural posterior estimation approach.

While our analysis of WD 1153+012 provides valuable insights, several uncertainties remain regarding the physics of hot DQ stars that would require a more rigorous approach. The Stark broadening treatment of carbon and oxygen lines in these unique atmospheres is still being refined, and the current models do not fully account for the effects of magnetism, which is prevalent in hot DQs and can significantly impact spectral features. A more comprehensive analysis would incorporate magnetic field modeling  and investigate the full UV-to-optical spectral range to better constrain the temperature structure and elemental abundances. It would be particularly insightful to apply the neural posterior technique with these updated physics considerations to the larger sample of 14 hot DQs analyzed in \citet{Dufour2007, Dufour2008} and \citet{Dufour2011}, especially those with high-quality MMT and Keck observations, to systematically constrain their physical parameters and potentially uncover evolutionary patterns within this rare spectral class.

\section{Conclusions}\label{sec:conca4}
In this work, we applied neural posterior estimation techniques for the physical parameter inference of white dwarfs from spectroscopic observations. We have demonstrated its ability to rapidly infer physical parameters for large numbers of DA and DB white dwarfs, as well as the more complex case of hot DQ stars with five parameters in total. Neural posterior estimation allowed to obtain the full posterior distribution of thousands of stars in a few seconds with a sampling speed of the order of 5000 samples per second, moving the overhead of inference from sampling the distribution for every object to training a single neural network per atmosphere type. Training the neural network takes between a few seconds to a minute, depending on the number of input dimensions and parameters to predict. The sampling speed is a significant increase from previous work for DA and DB \citep[500 samples per second using Random Forest interpolation][]{Chandra2020} or DZ white dwarfs \citep[12 to 24 hours per object to obtain the posterior][]{BadenasAgusti2024}. Furthermore, the parametric nature of neural posterior estimation can model any posterior distribution and be extremely useful for accurately measuring the parameters of white dwarfs with complex atmospheres. Increasing the number of parameters would not significantly affect inference time once the neural networks are trained, but would increase the data simulation time. In these scenarios, combining neural posterior estimation with machine learning interpolators to accelerate the sampling and training processes, respectively, would result in an extremely fast and robust methodology.

A key advantage of our approach over traditional methods is the ability to perform simulation-based calibration, as demonstrated in Section \ref{sec:valid}. This process, which would be prohibitively expensive with MCMC methods, allows us to ensure that the variance in the posterior distributions, and thus the reported uncertainties, are statistically sound. The TARP and SBC tests confirm that our posteriors are well-calibrated, with only a slight tendency toward under-confidence, providing reliable uncertainty quantification that is crucial for population studies. Furthermore, our methodology integrates seamlessly with photometric constraints, as shown in our iterative approach for carbon-atmosphere white dwarfs. This allows us to leverage complementary data sources while properly accounting for the systematic differences between spectroscopic and photometric analyses. The hybrid approach is particularly valuable for parameters like surface gravity in hot DQs, where line broadening physics introduces additional uncertainties.

Despite its advantages, our approach has several limitations worth addressing in future work. The current implementation requires automated mode identification for handling multimodal posteriors, which could benefit from more sophisticated techniques. Adding normality checks with KS tests would help identify when to switch methods for particularly complex distributions. Input spectra must match exactly the wavelength coverage used during training, limiting flexibility with heterogeneous datasets. Additionally, the photometric fitting component now represents the main computational bottleneck rather than the spectroscopic inference. Training neural posterior estimators for inference from photometric data would be possible and has been done in previous work for galaxy spectral energy distribution modelling \citep{Hahn2022, Khullar2022, Wang2023}, though it would be much more sensitive to issues with individual data points like missing observations for a single band.

Looking forward, this methodology presents exciting opportunities for extending to other challenging white dwarf subclasses, including magnetic and heavily metal-polluted cool white dwarfs. The computational efficiency of neural posterior estimation makes it ideal for upcoming surveys like SDSS-V, DESI, 4MOST, and eventually LSST, which will increase the known white dwarf population by over an order of magnitude.

\section*{Data Availability}
The Sloan Digital Sky Survey (SDSS DR17) optical spectra analysed in this work are publicly available from the SDSS Science Archive Server (\url{https://www.sdss.org}).
Gaia EDR3 astrometry and photometry are accessible through the ESA Gaia Archive (\url{https://gea.esac.esa.int/archive/}). All other data underlying this article will be shared on reasonable request to the corresponding author.

\section*{Acknowledgements}
We are grateful to the anonymous referee for a careful reading of our manuscript and for several constructive comments that helped to improve this paper. This work is supported in part by NSERC Canada, the Fund FRQ-NT (Qu\'ebec), and the Centre de recherche en astrophysique du Qu\'ebec (CRAQ). This work has made use of the sbi toolkit \citep{TejeroCantero2020}. This work presents results from the European Space Agency (ESA) space mission \emph{Gaia} and Sloan Digital Sky Survey. \emph{Gaia} data are being processed by the \emph{Gaia} Data Processing and Analysis Consortium (DPAC). Funding for the DPAC is provided by national institutions, in particular, the institutions participating in the \emph{Gaia} MultiLateral Agreement (MLA). Funding for the Sloan Digital Sky Survey (\url{https://www.sdss.org}) has been provided by the Alfred P. Sloan Foundation, the U.S. Department of Energy Office of Science, and the Participating Institutions. SDSS-IV acknowledges support and resources from the Center for High-Performance Computing at the University of Utah, and is managed by the Astrophysical Research Consortium for the Participating Institutions of the SDSS Collaboration. This research has also made use of the NASA Astrophysics Data System Bibliographic Services.



\bibliographystyle{mnras}
\bibliography{ms.bib}



\bsp	
\label{lastpage}
\end{document}